\documentclass[pra,
amssymb,amsmath,amsmath,showpacs,reprint,twocolumn]{revtex4-1}
\usepackage{color}
\usepackage{graphicx}
\usepackage{epstopdf}
\usepackage{natbib}
\usepackage{wasysym}
\usepackage{hyperref}
\usepackage{dcolumn}

\hypersetup{%
   pdfpagemode=None, %FullScreen,
   pdfstartpage=1,
   pdfstartview=FitH,
   pdfmenubar=true,
   pdftoolbar=true,
   colorlinks = true,
   linkcolor=blue,
   citecolor=blue,
   bookmarksopen=false
}

\newcommand{\Fkt}[1]{\,\mathsf {#1}}
\def\half{ \frac{1}{2}}
\def\openone{\leavevmode\hbox{\small1\kern-3.3pt\normalsize1}}

\ifx\Tr\renewcommand{\Tr}{\Fkt{Tr}} %has to commented out for IOP
\else\newcommand{\Tr}{\Fkt{Tr}}
\fi

\usepackage{booktabs}
\usepackage{multirow}
\usepackage{graphicx}
\usepackage{amsmath}
\usepackage{indentfirst}
\newcommand{\wigner}[6]{ \begin{pmatrix}
  #1 & #2 & #3 \\
  #4 & #5 & #6 
 \end{pmatrix}}

 \newcommand{\cpp}{\hat{U}_{\mbox{\tiny CPP}}}

\begin{document}
\title{Combining Slater-type orbitals and effective core potentials}

\author{\sc Micha\l\ Lesiuk}
\email{e-mail: lesiuk@tiger.chem.uw.edu.pl}
\author{\sc Aleksandra M. Tucholska}
\author{\sc Robert Moszynski}
\affiliation{\sl Faculty of Chemistry, University of Warsaw\\
Pasteura 1, 02-093 Warsaw, Poland}
\date{\today}

\begin{abstract}
We present a general methodology to evaluate matrix elements of the effective core potentials (ECPs) within 
one-electron basis set of Slater-type orbitals (STOs). The scheme is based on translation of individual STO
distributions in the framework of Barnett-Coulson method. We discuss different types of integrals which
naturally appear and reduce them to few basic quantities which can be calculated recursively or purely numerically.
Additionally, we consider evaluation of the STOs matrix elements involving the core polarisation potentials (CPP) and
effective spin-orbit potentials. Construction of the STOs basis sets designed specifically for use
with ECPs is discussed and differences in comparison with all-electron basis sets are briefly summarised. 
We verify the validity of the present approach by calculating excitation energies, static dipole polarisabilities and
valence orbital energies for the alkaline earth metals (Ca, Sr, Ba). Finally, we evaluate interaction energies,
permanent dipole moments and ionisation energies for barium and strontium hydrides, and compare them with the best
available experimental and theoretical data.
\end{abstract}

\maketitle

\section{Introduction}
\label{sec:intro}
In the last 40 years, Gaussian-type orbitals \cite{boys50,boys56} (GTOs) have undeniably taken the role of the routine
one-electron basis set for \emph{ab initio} calculations in molecular physics and quantum chemistry. Nonetheless, a
considerable interest has remained in the field of Slater-type orbitals (STOs) \cite{slater30,slater32} or more general
exponential-type orbitals (ETOs) \cite{homeier91,shull59}. This is motivated mainly by the superior analytical
properties of STOs (\emph{i.e.}, fulfilment of the nuclear cusp condition \cite{kato57} and correct
long-range decay \cite{agmon82,cycon87}) and their formal simplicity. 

The biggest obstacle connected with use of STOs is calculation of many-centre two-electron integrals which are
unavoidable in any molecular study. Interestingly enough, there has been a
significant progress on this issue in recent years. In fact, looking at only the past 15 years, there are many notable
works of Bouferguene \emph{et al.} \cite{bouferguene05a,bouferguene05b,bouferguene07,bouferguene09}, Rico \emph{et al.}
\cite{steinborn00,rico00,rico01a,rico01b,rico04a,rico04b,rico05,rico06,rico08a,rico08b,rico13}, Hoggan \emph{et al.}
\cite{hoggan01,hoggan02,hoggan03,hoggan04,hoggan09,hoggan10}, Pachucki
\cite{pachucki09,pachucki12,pachucki13a,pachucki13b,pachucki16}, and others
\cite{guidotti03,budzinski04,weniger00,weniger02,weniger12,harris02,barnett00,avery04,avery14,winkler13,gebremedhin13,
vukovic10,lesiuk12,lesiuk14a,lesiuk14b,lesiuk15,lesiuk16}. In particular, for the diatomic systems STOs can be now used
routinely \cite{lesiuk15}.

State-of-the-art \emph{ab initio} electronic structure calculations are important for the new field at the
border of chemistry and physics - studies of ultracold molecules.  Experimental advances in laser cooling and trapping
of neutral atoms have opened a door for the formation of ultracold diatomic molecules by
photoassociation \cite{julienne06a}, magnetoassociation \cite{julienne06b}, and vibrational cooling \cite{demille10}
techniques. To interpret the experimental observations, \emph{ab initio} calculations of the potential energy curves and
coupling matrix elements between the electronic states are crucial. Somewhat surprisingly,
spectroscopic and collisional studies of ultracold molecules mostly involve molecules with heavy atoms. See, for
instance, Refs. \cite{skomorowski12a,skomorowski12b,mcguyer13,mcguyer15a,mcguyer15b,mcguyer16} for joint experimental
and theoretical studies of new spectroscopic features of the strontium molecule. Electronic structure calculations can
also be used to predict new schemes for the formation of ultracold diatomic molecules
\cite{moszynski03,tomza11,tomza13a,krych11,tomza13b,tomza14}. Accurate interatomic interaction
potentials are also of great importance in search for a new physics, see \emph{e.g.}, the work on the YbF
molecule which is used in measurements of the permanent electric dipole moment (EDM) of the electron \cite{hudson11},
and determination of the proton-electron mass ratio time variation \cite{zelevinsky08}. One can also point out
the work of Schwertweger \emph{et al.}
\cite{schwertweger11} on the Sr$_2$ molecule where time dependence of the fine structure constant is considered.
Other examples of physically important diatoms include RbYb
molecule \cite{nemitz09,micheli06,zuchowski13} (a promising candidate for quantum computing), BaH
\cite{lane15,tarallo16,moore16} (preparation of ultracold hydrogen atoms) and others.

It must be stressed that in a majority of the studies mentioned above, accurate first principles calculations were
fundamental in understanding and interpretation of the experimental data. In particular, computations of the potential
energy surfaces and the corresponding coupling matrix elements appear to be of prime importance. This is also the area
where the Slater-type orbitals are the most advantageous. 

Unfortunately, in accurate \emph{ab initio} calculations for heavy elements one typically encounters two additional
major problems. First, the number of occupied orbitals becomes fairly substantial. This, by necessity, calls for
extended basis sets with high angular momenta, increasing the overall cost of correlated electronic structure
calculations. The second obstacle is the relative importance of the relativistic effects; for heavier elements they are
of a similar magnitude (or larger) as the electron correlation contribution \cite{visscher96a,visscher96b}. Moreover,
additivity of the latter two effects for heavy atoms is at best questionable. \cite{visscher96a,visscher96b}

There are several approaches available in the literature to handle the aforementioned problems and most of them are
based on the Dirac-Coulomb(-Breit) equations \cite{dirac28,bethe75}. This is done, \emph{e.g.}, by constructing an
approximate four-component
spinor expanded in a kinetically balanced basis set \cite{dyall91,visscher94,jong98,styszynski00}, or by decoupling the
small and large components of the spinor, so that the equations take a familiar two-component form
\cite{hess85,wolf02,reiher04,reiher06,liu05,barysz02,kedziera04a,kedziera04b}. Another idea developed independently
relies on the so-called regular approximations \cite{lenthe93,lenthe96}. In this paper we consider the effective core
potential (ECP)
approach \cite{dolg12} which may be viewed as a slightly less rigorous method than the former ones. However, little
accuracy is typically sacrificed
(especially for weakly bound systems) and the calculations simplify to a great extent. 

The fundamental idea behind ECPs is that the inner core orbitals of heavy elements are inert and do not change
significantly in different chemical environments. Therefore, their influence on the valence space can be modelled with a
proper pseudopotential (PP) \cite{foot1} which is, by definition, universal for a given element. This leads to an
approximate two-component valence relativistic wavefunction, obtained as an eigenfunction of the \emph{valence
only} Hamiltonian. This approach has two unquestionable advantages.
First, the inner core orbitals are removed from explicit consideration, so that the size of the one-electron basis
set is considerably reduced. Second, the scalar relativistic effects can be straightforwardly included in the
pseudopotential (by a proper parametrisation).

The main goal of this work is to combine the methodology of effective core potentials with the one-electron basis set
of Slater-type orbitals. We propose a general method to evaluate all necessary matrix elements by using analytical or
seminumerical techniques. Efficiency of the proposed algorithm is sufficient to perform general large-scale
calculations. Further in the paper, we consider the so-called core polarisation potentials
\cite{fuentealba82,muller84a,muller84b} which rely on the assumption
that the core is additionally polarisable. This captures the first-order effects of the implicit
core-valence correlations and significantly improves the description when the large-core pseudopotentials are used.
We also briefly consider effective spin-orbit pseudopotentials \cite{cohen79,ermler81,pitzer88} which allow for
an approximate computation of the spin-orbit splittings and couplings. Finally, we present results of test calculations
for alkaline earth metals (Ca, Sr, Ba) and predict properties of the corresponding hydrides.

In the paper we rely on the known special functions to simplify the derivations and the final formulae.
Our convention for all special functions appearing below is the same as in Ref. \cite{stegun72}.

\section{Theory}
\label{sec:theory}

In this section we introduce some important formulae which are used further in the paper. This is necessary to introduce
the notation and specify precise meaning of several quantities. This short mathematical preface may be useful for
readers who are not entirely familiar with employed concepts.

\subsection{Slater-type orbitals and spatial translations}
\label{subsec:sto}

Slater-type orbitals (STOs) \cite{slater30,slater32} have the following general form
\begin{align}
\label{sto1}
\chi_{nlm}(\textbf{r};\beta)=r^{n-1} e^{-\beta r}\,Y_{lm}(\theta,\phi),
\end{align}
where $Y_{lm}$ are the spherical harmonics defined according to the Condon-Shortley phase, $n$, $l$ are nonnegative
integers
satisfying $n>l$, and $\beta>0$ is a real parameter. The orbitals defined above are not normalised; we find this
convention to be more robust for the purposes of the paper. In order to restore the proper unity normalisation Eq.
(\ref{sto1}) has to be multiplied by $S_n(\beta)=(2\beta)^{n+1/2}/\sqrt{(2n)!}$.

Throughout the paper we shall extensively use the translation method for STOs in order to shift them to a convenient
point in space. This is achieved with help of the famous Barnett-Coulson $\zeta$-function method
\cite{coulson37,barnett51,barnett63}. Translation of the $ns$ STOs is given by the following two-range formula
\begin{align}
\label{trns}
 r_b^{n-1} e^{-\beta r_b} = \sum_{m=0}^\infty \frac{2k+1}{2}\,P_m(\cos \theta_a)\,\zeta_{nm}(\beta,r_a;R),
\end{align}
where $P_m$ are the ordinary Legendre polynomials, $R$ is the distance between centres $a$ and $b$, and $\zeta_{nm}$ are
given by the integral representation
\begin{align}
 \zeta_{nm}(\beta,r_a;R) = \int_0^\pi d\theta_a\, \sin \theta_a\, P_m(\cos \theta_a)\, r_b^{n-1}\, e^{-\beta r_b}.
\end{align}
From now on, we drop the parentheses from the definition of the $\zeta$-function, \emph{i.e.}, it is assumed
that $\zeta_{nm}:=\zeta_{nm}(\beta,r_a;R)$ unless explicitly stated otherwise. 

The above formal definitions are not particularly useful in actual applications. Instead, the following recursive
relations provide a starting point for further developments
\begin{align}
\begin{split}
\label{zetarec1}
 \zeta_{n+2,m} &= \big( r_a^2+R^2 \big)\,\zeta_{nm}-\frac{2\,r_a\,R}{2m+1} \\
 &\times \Big[ m\,\zeta_{n,m-1} + (m+1)\,\zeta_{n,m+1}\Big],
\end{split}
\end{align}
and
\begin{align}
\begin{split}
\label{zetarec2}
 \zeta_{1m} = \frac{\beta\, r_a\,R}{2m+1}\Big[ \zeta_{0,m-1} -
\zeta_{0,m+1}\Big].
\end{split}
\end{align}
The last formula is not valid for $m=0$ and the explicit expression should be used instead
\begin{align}
\label{zeta10}
 \zeta_{10} = \beta\,r_a\,R \Big[ \zeta_{00} - \zeta_{01} \Big] + e^{-\beta(r_a+R)}.
\end{align}
To initiate the recursive process one requires the following starting values
\begin{align}
\label{zeta0k}
 \zeta_{0m} = \frac{2\beta}{\pi}\,i_m(\beta\, r_<)\,k_m(\beta\,r_>),
\end{align}
where $i_m$ and $k_m$ are the modified spherical Bessel functions of the first and second kind \cite{stegun72},
respectively, and $r_<=\min(r_a,R)$, $r_>=\max(r_a,R)$. For convenience of the reader, we gathered all properties of
the modified spherical Bessel functions which are important here in the Supplemental Material \cite{supplement}.
Equations in the Supplemental Material are referenced with prefix ``S'', \emph{e.g.}, the sixth equation in the
Supplemental Material is referenced as Eq. (S6). 

\begin{widetext}
In order to spatially shift STOs of the form (\ref{sto1}) one needs to combine Eq. (\ref{trns}) with the well-known
translation formula for the regular solid harmonics, Ref. \cite{arfken}, pp. 797. This leads to
\begin{align}
\label{trsto1}
\begin{split}
 \frac{1}{\sqrt{\pi}}\,r_b^{n-1}\,e^{-\beta r_b}\,Y_{lm}(\theta_b,\phi_b) &= (-1)^{l-m}\,\big(2l+1\big)
 \sum_{\lambda=0}^l \frac{r_a^\lambda}{\sqrt{2\lambda+1}}\,Y_{\lambda m}(\theta_a,\phi_a)\,(-R)^{l-\lambda}
 \wigner{\lambda}{l-\lambda}{\lambda}{m}{0}{-m} \\
 & \times \binom{2l}{2\lambda}^{1/2}
 \sum_{k=0}^\infty \sqrt{2k+1}\,Y_{k0}(\theta_a,\phi_a)\,\zeta_{n-l,k}(\beta,r_a;R),
\end{split}
\end{align}
with the usual notation for the Wigner $3J$ symbols, Ref. \cite{arfken}, pp. 270.
\end{widetext}

A short comment on the nature of the above expression is necessary. It is well-known that the strongest drawback of the
Barnett-Coulson method is that it leads, in general, to infinite series. Worse than that, these series tend to converge
extremely slowly; in some cases a logarithmic convergence pattern has been observed \cite{lowdin56,flygare66}. This
limits the applicability of the method significantly and forced some authors to apply convergence acceleration
techniques \cite{petersson67,bouferguene98}. Fortunately, this issue is absent in \emph{all} final formulae derived in
this paper. In most cases, the sum over $k$ truncates as a result of the triangle conditions for the Wigner $3J$
symbols, see Ref. \cite{arfken}, pp. 803.

Before the end of the present section we would like to point out that there exist some other methods for translation of
STOs, including one-range and two-range formulae, Refs.
\cite{danos65,filter80,homeier92,novosadov,steinborn80,weniger85,guseinov76,guseinov78,guseinov80,guseinov85,guseinov88}
, yet we have not found these alternative formulations to be particularly advantageous in the present case compared to
the standard Barnett-Coulson scheme, Eq. (\ref{trsto1}). General theory of addition theorems was given in a pedagogical
way by Weniger \cite{weniger00,weniger02}. Note that mathematical correctness (and usefulness) of some formulations of
the addition theorems is still subject to a debate \cite{weniger12}.

\subsection{Pseudopotentials parametrisation}
\label{subsec:param}

As already mentioned in the introduction, in calculations involving ECPs one considers the following valence-only
Hamiltonian \cite{dolg12}
\begin{align}
\label{hval}
\begin{split}
\hat{H}_v &= \sum_i^{n_v} \left[ -\half\nabla_i^2 + \sum_a \left[ -\frac{Q_a}{r_{ia}} + \hat{U}_{\mbox{\scriptsize
 PP}}^a(r_{ia}) \right]
\right]\\ 
&+ \sum_{i<j}^{n_v} \frac{1}{r_{ij}} + \sum_{ab} \frac{Q_a Q_b}{r_{ab}} + \cpp,
\end{split}
\end{align}
where $i,j,...$ denote the electrons, $a,b,...$ denote the nuclei, $\hat{U}_{\mbox{\scriptsize PP}}^a$ is the
pseudopotential of the core
$a$ with charge $Q_a$, and $n_v$ is the number of valence electrons. The term arising from the core polarisation
potential ($\cpp$) will be specified further in the text.

Let us briefly discuss the construction of the effective core potentials. They are divided into the spin-averaged and
spin-dependent terms, $U_{\mbox{\scriptsize PP}}^a = U_{\mbox{\scriptsize PP,av}}^a + U_{\mbox{\scriptsize PP,so}}^a$.
Typically, the first term is included explicitly in the electronic structure calculations whilst the second is treated
perturbatively. Both of these potentials are represented in a semi-local form
\begin{align}
\label{ecp1}
\begin{split}
\hat{U}_{\mbox{\scriptsize PP,av}}^a(r_{ia}) &= U_L^a(r_{ia}) 
+ \sum_{l=0}^{L-1}\sum_{m=-l}^{+l} |lm\rangle_a
\\ &\times\left[ U_l^a(r_{ia}) - U_L^a(r_{ia}) \right] 
\,_a\langle lm|,
\end{split}
\end{align}
and
\begin{align}
\label{ecpso1}
\begin{split}
\hat{U}_{\mbox{\scriptsize PP,so}}^a(r_{ia}) = \sum_{l=0}^{L-1}\sum_{m=-l}^{+l} \frac{2\,\Delta
U_l^a(r_{ia})}{2l+1}\,|lm\rangle_a\,\,
\mathbf{l}_{ia}\cdot\mathbf{s}_i\,\,_a\langle lm|,
\end{split}
\end{align}
where $L$ is the highest angular momentum of the orbitals in the core $a$, $\mathbf{l}_{ia}$ is the (orbital) angular
momentum operator corresponding to the centre $a$ and the electron $i$, $\mathbf{s}_i$ is the spin operator of the
electron $i$, and $\langle lm|_a$ are projection operators on the spherical harmonics $Y_{lm}$ placed at the centre $a$.
Presence of the projection operators assures that orbital components of different angular momenta connect with proper
radial functions. Parenthetically, it is observed that $U_l^a(r_{ia})$ are nearly identical for $l>L$ which justifies
the rearrangements in Eqs. (\ref{ecp1}) and (\ref{ecpso1}).

To specify a family of pseudopotentials a precise form of the radial components, $U_l^a(r)$, must be
given. It is very common to use a short linear combination of the radial Gaussian functions \cite{dolg12}
\begin{align}
\label{ecppar}
 r^2\,U_l^a(r) = \sum_k A_{kl}^a\,r^{n_{kl}} e^{-B_{kl}^a r^2},
\end{align}
where $n_{kl}$, $A_{kl}^a$ and $B_{kl}^a$ are adjustable parameters. Their determination for a given atom is far 
from trivial and strategies of the so-called energy-consistent \cite{preuss55,topp73,flad79,preuss81}, shape-consistent
\cite{phillips59,kahn76,baran88,wang00}, and other \cite{dolg12} pseudopotentials were developed.

\subsection{Effects of the core polarisation}
\label{subsec:cpptheo}

The so-called core polarisation potentials \cite{muller84a,muller84b,fuentealba82} (CPP) constitute a method to improve
upon the approximations underlying the ordinary ECPs. The core is allowed to be polarisable, \emph{i.e.}, reorientation
of valence electrons in a molecular environment creates an induced dipole moment of the core. By simple electrostatic
arguments, the value of this dipole moment is assumed to be proportional to the strength of the electric field at the
core. This gives rise to the total potential $\cpp=\sum_a \cpp^a$ in the form 
\begin{align}
 \label{cpp1}
 \cpp^a = \cpp^{[0],a} + \cpp^{[1],a} + \cpp^{[2],a},
\end{align}
\begin{align}
 \label{cpp2}
 \hat{U}_{cpp}^{[0],a} = -\half \alpha_a \sum_{b,c\neq a} Q_b Q_c \frac{\mathbf{R}_{ab}\cdot\mathbf{R}_{ac}}{R_{ab}^3
 R_{ac}^3},
\end{align}
\begin{align}
\begin{split}
 \label{cpp3}
 \hat{U}_{cpp}^{[1],a} &= -\half \alpha_a \sum_i \frac{1}{r_{ia}^4} C^2\big(r_{ia},\delta_a\big) \\
 &+ \alpha_a \sum_i \sum_{b\neq a} Q_b\,\frac{\mathbf{r}_{ia}\cdot \mathbf{R}_{ab}}{r_{ia}^3 R_{ab}^3}
 \,C(r_{ia},\delta_a),
\end{split}
\end{align}
\begin{align}
 \label{cpp4}
 \hat{U}_{cpp}^{[2],a} = -\alpha_a \sum_{j<i} \frac{\mathbf{r}_{ia}\cdot \mathbf{r}_{ja}}{r_{ia}^3 r_{ja}^3}
 \,C(r_{ia},\delta_a)\,C(r_{ja},\delta_a),
\end{align}
where the consecutive terms are the scalar, one- and two-electron components. In the above expression $\alpha_a$ is the
polarisability of the core $a$, determined from separate theoretical calculations or by semiempirical adjustment;
$C(r_{ia},\delta_a)$ is the cutoff function assuring that the potential is regular when electron $i$ is at the core $a$.
The form of the cutoff function as well as the value of the cutoff parameter $\delta_a$ are arbitrary. 
The following expression is frequently used
\begin{align}
\label{cut}
 C(r,\delta) = \Big(1-e^{-\delta r^2}\Big)^{\bar{n}},
\end{align}
where $\bar{n}$ is either $1$ (Stoll-Fuentealba \cite{fuentealba82}) or $2$ (M\"{u}ller-Meyer
\cite{muller84a,muller84b}). The optimal values of $\delta_a$ are determined by numerical experimentation for each atom
separately.

\subsection{Basic integrals}
\label{subsec:basicint}

Further in the text we show that all matrix elements involving averaged and spin-orbit pseudopotentials can be
expressed through the following family of one-dimensional integrals
\begin{align}
 \label{f000}
 &\mathcal{F}^{\,0}_n\big(x,y\big) = \int_0^\infty dr\;r^n\,e^{-x r- y r^2},\\
 \label{fbig}
 &\mathcal{F}^{\,>}_n\big(x,y\big) = \int_1^\infty dr\;r^n\,e^{-x r- y r^2},\\
 \label{fsmall}
 &\mathcal{F}^{\,<}_n\big(x,y\big) = \int_0^1 dr\;r^n\,e^{-x r- y r^2}.
\end{align}
For evaluation of the core polarisation potentials matrix elements one additionally requires integrals
with a logarithmic (albeit integrable) singularity, \emph{e.g.}
\begin{align}
 \label{g000}
 \mathcal{G}^{\,0}_n\big(x,y\big) = \int_0^\infty dr\;\ln r\,r^n\,e^{-x r- y r^2},
\end{align}
and similarly for $\mathcal{G}^{\,>}_n$ and $\mathcal{G}^{\,<}_n$.

The issue of calculation of the above integrals is fairly technical and marred with numerical problems. For
completeness, in the Supplemental Material we present an exhaustive description of the methods we recommend for
calculation of these basic quantities. Additionally, a special case of these integrals were considered in Ref.
\cite{silkowski15}. Note that Eqs. (\ref{f000}) and (\ref{fsmall}) are well-defined only for $n\geq0$; this
restriction does not hold for Eq. (\ref{fbig}).

\section{Spin-averaged and spin-orbit matrix elements}
\label{sec:matrix}

In the remainder of the paper we consider calculation of ECPs matrix elements for diatomic systems only.
This is mainly because the issue of exact calculation of the two-electron integrals for polyatomic molecules has not
been fully resolved yet. Consequently, we adopt a coordinate system where both atoms are located on the $z$ axis and
separated by a distance $R$. 

For calculations on general polyatomic systems one needs the following matrix elements involving the spin-averaged
potentials
\begin{align}
 \label{ibca1}
 I_{\mbox{\scriptsize bca}} = \langle 
 \chi_{n_bl_bm_b}(\textbf{r}_b;\beta_b)|\,\hat{U}_{\mbox{\scriptsize
 PP,av}}^c(r_{c})|\,\chi_{n_al_am_a}(\textbf{r}_a;\beta_a)\rangle.
\end{align}
By using Eq. (\ref{ecp1}) and after simple manipulations one can rewrite the above expression as
\begin{align}
 I_{\mbox{\scriptsize bca}} = I_{\mbox{\scriptsize bca}}^{\mbox{\scriptsize
loc}}+I_{\mbox{\scriptsize bca}}^{\mbox{\scriptsize nloc}}
\end{align}
where
\begin{align}
 \label{ibca2loc}
\begin{split}
 &I_{\mbox{\scriptsize bca}}^{\mbox{\scriptsize loc}} = 
 \langle\chi_{n_bl_bm_b}(\textbf{r}_b;\beta_b)|\,U_L^c(r_{c})\,|\chi_{n_al_am_a}(\textbf{r}_a;\beta_a)\rangle,
\end{split}
\end{align}
and
\begin{align}
 \label{ibca2nloc}
\begin{split}
 &I_{\mbox{\scriptsize bca}}^{\mbox{\scriptsize nloc}} = 
 \sum_{KM}^{L-1} \int_0^\infty dr_c\,r_c^2\;\langle \chi_{n_bl_bm_b}(\textbf{r}_b;\beta_b)|KM\rangle_c \\
 &\times\left[ U_K^c(r_{c}) - U_L^c(r_{c}) \right] \,_c\langle KM|\chi_{n_al_am_a}(\textbf{r}_a;\beta_a)\rangle.
\end{split}
\end{align}
The approach adopted here depends on the relative location of the centres.

The main difficulty connected with the calculation of the pseudopotentials matrix elements is the presence of the
projection operators if the orbitals are placed on different centres. Additionally, pseudopotentials are typically
parametrised in terms of the Gaussian-type expansions, Eq. (\ref{ecppar}), which leads to mixed Slater-Gaussian type
basic integrals. The latter are usually not easily expressible through the standard elementary and special functions,
and new techniques need to be developed to handle them.

\subsection{Spin-averaged potentials, $I_{\mbox{\scriptsize aaa}}$ type}
\label{subsec:ecpaaa}

Let us first consider the atomic case, $a=b=c$. Due to orthogonality of the spherical harmonics the matrix element
simplifies to
\begin{align}
\label{ieaaa}
\begin{split}
 &I_{\mbox{\scriptsize aaa}} = I_{\mbox{\scriptsize aaa}}^{\mbox{\scriptsize
loc}}+I_{\mbox{\scriptsize aaa}}^{\mbox{\scriptsize nloc}} = \\
 &\begin{cases}
 \langle\chi_{n_a'l_a'm}(\textbf{r}_a;\beta_a')|\,U_L^a(r_{a})\,|\chi_{n_al_am}(\textbf{r}_a;\beta_a)\rangle, &
l_a\geq L \\ 
 \langle\chi_{n_a'l_a'm}(\textbf{r}_a;\beta_a')|\,U_{l_a}^a(r_{a})\,|\chi_{n_al_am}(\textbf{r}_a;\beta_a)\rangle, &
l_a<L,
 \end{cases}
\end{split}
\end{align}
provided that $l_a=l_a'$ and $m_a=m_a'=m$. Otherwise, the result vanishes due to the spherical symmetry of the
integrand. Evaluation of the remaining integrals is now elementary; making use of Eqs. (\ref{sto1}) and (\ref{ecppar})
\begin{align}
\begin{split}
 &I_{\mbox{\scriptsize aaa}}=
 \sum_k A_{kl}^a\,\mathcal{F}^{\,0}_{n_a+n_a'+n_{kl}-2}\Big(\beta_a+\beta_a',B_{kl}^a\Big),
\end{split}
\end{align}
where $\mathcal{F}^{\,0}_n\big(x,y\big)$ is defined by Eq. (\ref{f000}).

\subsection{Spin-averaged potentials, $I_{\mbox{\scriptsize baa}}$ type}
\label{subsec:ecpbaa}

Let us now consider the first of the two-centre matrix elements, $I_{\mbox{\scriptsize baa}}$. One can easily see that
they obey formally the same expression as (\ref{ieaaa}), but $l_a$ and $l_b$ do not need to be equal. However, the
requirement $m_a=m_b$ still holds as a consequence of the axial symmetry. Translating STO from the centre $b$ to the
point $a$
\begin{align}
\begin{split}
&\frac{(-1)^{l_b}}{\sqrt{2l_a+1}}\langle\chi_{n_bl_bm}(\textbf{r}_b;\beta_b)|\,U_L^a(r_{a})\,
|\chi_{ n_al_am}(\textbf{r} _a;\beta_a)\rangle = \\
&\frac{2l_b+1}{2}\sum_{\lambda=0}^{l_b} \binom{2l_b}{2\lambda}^{1/2}
\wigner{\lambda}{l_b-\lambda}{l_b}{m}{0}{-m}\left(-R\right)^{l_b-\lambda}\mathcal{A}_\lambda,
\end{split}
\end{align}
where
\begin{align}
\begin{split}
&\mathcal{A}_\lambda = \sum_k \big(2k+1\big) \wigner{\lambda}{k}{l_a}{0}{0}{0}
\wigner{\lambda}{k}{l_a}{m}{0}{-m} \\
&\times \sum_m A_{mL}^a\,W_{\lambda+n_a+n_{mL}-1,n_b-l_b,k}\big(\beta_a,\beta_b,B_{mL};R\big).
\end{split}
\end{align}
The remaining one-dimensional integration is confined to the following formula
\begin{align}
\label{wlmn}
\begin{split}
W_{lmn}\big(\alpha,\beta,\gamma;R\big) = 
\int_0^\infty dr\; r^l\,\zeta_{mn}\big(\beta,r;R)\,e^{-\alpha r-\gamma r^2}.
\end{split}
\end{align}

A straightforward approach to Eq. (\ref{wlmn}) is to use a quadrature of some sort and
treat the integrals in a purely numerical fashion. However, the integrand possesses a derivative discontinuity
(\emph{i.e.}, a cusp) at $r=R$. This makes integration with standard Gaussian quadratures difficult. For a reasonable
performance one would need to divide the integration range into two subintervals, $\big[0,R\big]$ and
$\big[R,\infty\big]$, and treat each of them separately, possibly with different integration rules. This, in turn,
increases the computational costs as the integration needs to be performed for a large set of parameters $l$, $m$, $n$.

An alternative approach which we propose here relies on the recursive relations (\ref{zetarec1}) and (\ref{zetarec2}).
By inserting them into the definition (\ref{wlmn}) one arrives at
\begin{align}
\begin{split}
 W_{l,m+2,n} &= W_{l+2,mn} + R^2\,W_{lmn} - \frac{2 R}{2n+1} \\
 &\times \Big[ n\,W_{l+1,m,n-1} + \big(n+1\big) W_{l+1,m,n+1} \Big],
\end{split}\\
 W_{l1n} &= \frac{\beta R}{2n+1} \Big[ W_{l+1,0,n-1} - W_{l+1,0,n+1} \Big],
\end{align}
and
\begin{align}
\begin{split}
 W_{l10} &= \beta R \Big[ W_{l+1,00}-W_{l+1,01} \Big] \\
 &+ e^{-\beta R} \mathcal{F}^{\,0}_l\big(\alpha+\beta,\gamma\big).
\end{split}
\end{align}
To initiate the above recursions one needs $W_{l0n}$ which can be expressed with help of Eq. (\ref{zeta0k}) as
\begin{align}
\begin{split}
 W_{l0n} &= \frac{2\beta}{\pi} R^{l+1} \Big[ k_n\big(\beta R\big)\,\mathcal{I}_{ln}\big(\alpha R,\beta R,\gamma R\big)
\\ &+ i_n\big(\beta R\big)\,\mathcal{K}_{ln}\big(\alpha R,\beta R,\gamma R\big) \Big],
\end{split}
\end{align}
where
\begin{align}
\label{defbigi}
&\mathcal{I}_{ln}\big(\alpha,\beta,\gamma\big) = \int_0^1 dx\; x^l\, i_n(\beta x)\, e^{-\alpha x-\gamma x^2}, \\
\label{defbigk}
&\mathcal{K}_{ln}\big(\alpha,\beta,\gamma\big) = \int_1^\infty dx\; x^l\, k_n(\beta x)\, e^{-\alpha x-\gamma x^2}.
\end{align}
The latter two integrals can be integrated numerically to a very good precision. This approach is reasonable if one does
not care about the timings of the calculations (\emph{e.g.}, for benchmark purposes). However, to reach a
computational cost comparable with GTOs and use ECPs efficiently for large systems better procedures are required,
preferably recursive. They are described in detail in next paragraphs.

Let us begin with the first class of integrals, $\mathcal{I}_{ln}$ (we drop the parentheses from now on). By using the
relation (S5) one arrives at
\begin{align}
\label{bigirec}
 \mathcal{I}_{l,n-1} = \frac{2n+1}{\beta}\, \mathcal{I}_{l-1,n} + \mathcal{I}_{l,n+1}.
\end{align}
This recursion needs to be carried out in the direction of decreasing $n$ in order to maintain the numerical
stability. To start the process (\ref{bigirec}) one requires $\mathcal{I}_{lN}$ for two neighbouring (large) $N$ and
$\mathcal{I}_{0n}$. We propose to evaluate both of them by inserting the power series expansion of $i_n(x)$ around the
origin into the definition (\ref{defbigi})
\begin{align}
\label{ilnpow}
\mathcal{I}_{ln} = \beta^n \sum_{k=0}^\infty \frac{\Big(\half \beta^2\Big)^k}{k!\,(2n+2k+1)!!}
\,\mathcal{F}^{\,<}_{n+l+2k}\big(\alpha,\gamma\big).
\end{align}
Since the above summation is infinite and for practical reasons needs to be truncated, it is helpful to
estimate in advance how many terms are required to achieve convergence.

We first note that the rate of convergence of Eq. (\ref{ilnpow}) is not significantly affected by a change of values 
$l$, $\alpha$, and $\gamma$, the only important variables being $\beta$, $n$. The sum (\ref{ilnpow}) converges faster
when $\beta$ decreases or $n$ increases. Therefore, we can consider the worst-case scenario of $\mathcal{I}_{00}$ as a
function of $\beta$. Making use of the relationship $\mathcal{F}^{\,<}_n\big(\alpha,\gamma\big) \leq \frac{1}{n+1}$ one
arrives at the formal upper bound
\begin{align}
\label{i00}
\mathcal{I}_{00} \leq \sum_{k=0}^\infty \frac{\beta^{2k}}{(2k)!(2k+1)^2}.
\end{align}
One can assume that the convergence pattern of the above series is very similar to the original $\mathcal{I}_{00}$.
The number of terms necessary to achieve convergence for a given $\beta$ can be estimated by solving the
equality $\beta^{2k}=\epsilon\,(2k)!\,(2k+1)^2$ and rounding up to the closest integer value ($\epsilon$ is the
prescribed accuracy goal). We obtained numerical solutions of Eq. (\ref{i00}) for a finite set of $\beta$ and fitted
them with a linear function, giving $n_{\mbox{\scriptsize terms}} = 0.68 \beta +
29.5$. This estimation is reliable for all $\beta$, but it tends to overshoot $n_{\mbox{\scriptsize terms}}$ slightly,
especially for smaller $\beta$.

The method based on the infinite summation is quite successful for small and moderate $\beta$ but becomes tedious when
the values of $\beta$ get large. It typically occurs for stretched molecules or for extended basis sets with high 
exponents. To avoid laborious summations in such situations we present a large $\beta$ asymptotic expansion of
the functions $\mathcal{I}_{ln}$. The derivation begins by rewriting Eq. (\ref{defbigi}) as a difference of two
integrals over the intervals $[0,\infty]$ and $[1,\infty]$. In the first integral one needs to exchange
the variables to $\beta x$ and subsequently expand the Gaussian function under the integral sign in power series. The
remaining integral can be recognised as the Legendre function of the second kind $Q_n$ by means of the analytic
continuation. This finally leads to the asymptotic formula for the first part
\begin{align}
\label{asymi1}
\int_0^\infty x^l\,i_n(\beta x)\,e^{-\alpha x - \gamma x^2} \sim \frac{1}{\beta^{l+1}} \sum_{k=0}^\infty
\left(\frac{\gamma}{\beta^2}\right)^k Q_n^{(l)}\left(\frac{\alpha}{\beta}\right),
\end{align}
where the subscript in $Q_n$ denotes differentiation with respect to the main argument. Calculation of the Legendre
functions and their derivatives is a standard task as has been discussed many times in the literature
\cite{olver83,gil98,schneider10}. Let us
pass to the second part, \emph{i.e.}, the integral over $[1,\infty]$. Note that in this integral (contrary to the
former) the argument of the Bessel function is always large for large $\beta$. Therefore, we can use the large-argument
expansion of the Bessel function given by Eq. (S7) in Supplemental Material. This straightforwardly leads
to the formula
\begin{align}
\label{asymi2}
\int_1^\infty x^l\,i_n(\beta x)\,e^{-\alpha x - \gamma x^2} \sim \half \sum_{k=0}^\infty \frac{a_{kn}}{\beta^{k+1}}
\mathcal{F}^{\,>}_{l-k-1}\big(\alpha,\gamma\big).
\end{align}
By combining Eqs. (\ref{asymi1}) and (\ref{asymi2}) one obtains the final large-$\beta$ asymptotic
expansion of the integrals (\ref{defbigi}). Explicit expressions for the coefficients in the above expression are given
in Supplemental Material. 

Passing to the second class of integrals, $\mathcal{K}_{ln}$, and inserting the explicit formula for $k_n(z)$, Eq.
(S3), leads to
\begin{align}
\label{klnex}
\mathcal{K}_{ln} = \frac{\pi}{2\beta}\sum_{k=0}^n\frac{(n+k)!}{(2\beta)^k\,k!\,(n-k)!}\,
\mathcal{F}^{\,>}_{l-k-1}\big(\alpha+\beta,\gamma\big),
\end{align}
which makes the evaluation elementary. Note that all terms in the above sum are strictly positive, so that no
cancellations are possible and the final result acquires the same accuracy as the supplied values of ${F}^{\,>}_l$.

\subsection{Spin-averaged potentials, $I_{\mbox{\scriptsize bab}}$ type}
\label{subsec:ecpbab}

In the case of $I_{\mbox{\scriptsize bab}}$ configuration there are no simplifications analogous as in Eq. (\ref{ieaaa})
and we must use Eqs. (\ref{ibca2loc}) and (\ref{ibca2nloc}) as they stand. Therefore, the local and nonlocal parts need
to be treated separately in this case. Considering the local part, note that in Eq. (\ref{ibca2loc}) both STOs occupy
the same centre. Therefore, one can expand the product of two STOs into a linear combination of STOs by using standard
relations for coupling of the angular momenta. As a result, the integrals $I_{\mbox{\scriptsize bab}}^{\mbox{\scriptsize
loc}}$ can easily be expressed in terms of 
\begin{align}
\label{babloc}
\begin{split}
&\int d\mathbf{r}\;r_b^{n_{ab}-1} e^{-\beta_{ab} r_b}
\;Y_{l_{ab},0}(\theta_b,\phi_b)\,U_L^a(r_a)= \\
& (-1)^{l_{ab}}\frac{2l_{ab}+1}{\sqrt{\pi}}
\sum_\lambda^{l_{ab}} \wigner{\lambda}{l_{ab}-\lambda}{l_{ab}}{0}{0}{0} \binom{2l_{ab}}{2\lambda}^{1/2}\\
&(-R)^{l_{ab}-\lambda}\sum_k A_{kl}^a\,W_{\lambda+n_{kl},n_{ab}-\lambda,\lambda}\big(0,\beta_{ab},B_{kl}^a;R\big) 
\end{split}
\end{align}
where we have made use of Eq. (\ref{trsto1}) and integrated over the angles. In the above expression a handful of
quantities has been introduced, \emph{i.e.}, $n_{ab}=n_a+n_b-1$, $\beta_{ab}=\beta_a+\beta_b$, and $|l_a-l_b|\leq
l_{ab}\leq l_a+l_b$ (note that $n_{ab}>l_{ab}$).

The remaining one-dimensional integrals in Eq. (\ref{babloc}) are of the same class as defined by Eq. (\ref{wlmn})
but with $\alpha=0$. Theoretically, this brings a degree of simplification and allows for a more robust scheme. However,
we found that it is not worth increasing the size and complexity of the code by including separate
routines for the case $\alpha=0$. Therefore, we recommend that the case $\alpha=0$ is treated with general techniques
described above. There are no singularities or numerical instabilities in these expressions as $\alpha$ approaches zero,
so that the codes can be reused with no changes.

Let us now consider the calculation of the nonlocal term, $I_{\mbox{\scriptsize bab}}^{\mbox{\scriptsize nloc}}$. This
case is much more troublesome due to the fact that the coupling of the angular momenta cannot be used before the
translation of the orbitals. Therefore, both STOs need to be shifted independently from the centre $b$ to the centre
$a$. After some algebra one finds
\begin{align}
\begin{split}
 I_{\mbox{\scriptsize bab}}^{\mbox{\scriptsize nloc}} &= (-1)^{l_a+l_b} \frac{(2l_a+1)(2l_b+1)}{4} 
\sum_{K=0}^{L-1} (2K+1)\,\mathcal{A}_K,
\end{split}
\end{align}
where
\begin{align}
\begin{split}
&\mathcal{A}_K = \sum_{\lambda_a}^{l_a} (-R)^{l_a-\lambda_a}  \wigner{\lambda_a}{l_a-\lambda_a}{l_a}{m}{0}{-m}
\binom{2l_a}{2\lambda_a}^{\half} \\ 
&\sum_{k_a}(2k_a+1) \wigner{K}{k_a}{\lambda_a}{0}{0}{0} \wigner{K}{k_a}{\lambda_a}{-m}{0}{m}
\mathcal{B}_{l_a m_a},
\end{split}
\end{align}
and analogously
\begin{align}
\begin{split}
&\mathcal{B}_{l_a m_a} = \sum_{\lambda_b}^{l_b} (-R)^{l_b-\lambda_b}  \wigner{\lambda_b}{l_b-\lambda_b}{l_b}{m}{0}{-m}
\binom{2l_b}{2\lambda_b}^{\half} \\ 
&\sum_{k_b}(2k_b+1) \wigner{K}{k_b}{\lambda_b}{0}{0}{0} \wigner{K}{k_b}{\lambda_b}{-m}{0}{m} \\
&\sum_p A_{pL}^a\,U_{\lambda_a+\lambda_b+n_{pL},k_a,k_b}^{n_a-l_a,n_b-l_b}(\beta_b,\beta_b',B_{pL}^a;R).
\end{split}
\end{align}
Finally, the innermost integral can be expressed as
\begin{align}
\label{udef}
\begin{split}
 &U_{l\,n_1n_2}^{m_1m_2}(\beta_1,\beta_2,\gamma;R) = \\
&\int_0^\infty dr\,r^l\,\zeta_{m_1n_1}(\beta_1,r;R)\,\zeta_{m_2n_2}(\beta_2,r;R)e^{-\gamma r^2},
\end{split}
\end{align}
reducing all matrix elements to a definite one-dimensional integration. Let us note at this point that the integrals
Eq. (\ref{udef}) are invariant with respect to permutation $1\leftrightarrow2$ of all indices (including the
nonlinear parameters), \emph{i.e.},
$U_{l\,n_1n_2}^{m_1m_2}(\beta_1,\beta_2,\gamma;R)=U_{l\,n_2n_1}^{m_2m_1}(\beta_2,\beta_1,\gamma;R)$.

Clearly, the integrals $U_{l\,n_1n_2}^{m_1m_2}$ are the most complicated quantities appearing in the theory. Since they
are five-index objects, any numerical integration is expected to be prohibitively expensive. Therefore,
the recursive techniques are preferred despite the necessity to operate in many dimensions. Derivation of
the recursive formulae for the basic integrals $U_{l\,n_1n_2}^{m_1m_2}$ follows along a line similar as in the previous
subsection. Let us insert Eqs. (\ref{zetarec1}) and (\ref{zetarec2}) into the definition of $U_{l\,n_1n_2}^{m_1m_2}$.
After some rearrangements one obtains
\begin{align}
\label{urec1}
\begin{split}
U_{l\,n_1n_2}^{m_1+2,m_2} &= U_{l+2,n_1n_2}^{m_1m_2} + R^2\,U_{l\,n_1n_2}^{m_1m_2}
-\frac{2R}{2n_1+1} \\
&\times \Big[ n_1\,U_{l\,n_1-1,n_2}^{m_1m_2} + (n_1+1)\,U_{l\,n_1+1,n_2}^{m_1m_2} \Big],
\end{split}
\end{align}
and
\begin{align}
\label{urec2}
 U_{l\,n_1n_2}^{1,m_2} = \frac{\beta_1 R}{2n_1+1} \Big[
 U_{l+1,n_1-1,n_2}^{0,m_2} - U_{l+1,n_1+1,n_2}^{0,m_2} \Big].
\end{align}
The exceptions from the above relation are the integrals with $n_1=0$ which have to be calculated according to
Eq. (\ref{zeta10}) instead
\begin{align}
\label{urec3}
\begin{split}
 U_{l\,0,n_2}^{1,m_2} &= \beta_1 R \Big[ U_{l+1,0,n_2}^{0,m_2} - U_{l+1,1,n_2}^{0,m_2} \Big] \\
 &+ e^{-\beta_1 R}\,W_{l,m_2,n_2}(\beta_1,\beta_2,\gamma;R).
\end{split}
\end{align}
The recursion relations which allow to increase the second pair of indices can be obtained by using the aforementioned
symmetry property. 

The above relations allow to calculate $U_{l\,n_1n_2}^{m_1m_2}$ with nonzero $m_1$, $m_2$ starting solely with the
integrals $U_{l\,n_1n_2}^{00}$. The latter obey the relationship
\begin{align}
\begin{split}
 &U_{l\,n_1n_2}^{00} = \frac{4\beta_1\beta_2}{\pi^2} R^{l+1} \\
 &\times\Big[ k_{n_1}(\beta_1 R)\,k_{n_2}(\beta_2 R)\,J_{n_1n_2}^l(\beta_1 R,\beta_2 R,\gamma R^2) \\
 &+ i_{n_1}(\beta_1 R)\,i_{n_2}(\beta_2 R)\,M_{n_1n_2}^l(\beta_1 R,\beta_2 R,\gamma R^2) \Big],
\end{split}
\end{align}
where
\begin{align}
\label{jn1n2}
 \mathcal{J}_{n_1n_2}^l(\beta_1,\beta_2,\gamma) &= \int_0^1 dx\,x^l\,i_{n_1}(\beta_1 x)\,i_{n_2}(\beta_2 x)\,e^{-\gamma
x^2},\\
\label{kn1n2} 
 \mathcal{K}_{n_1n_2}^l(\beta_1,\beta_2,\gamma) &= \int_1^\infty dx\,x^l\,k_{n_1}(\beta_1 x)\,k_{n_2}(\beta_2
x)\,e^{-\gamma x^2},
\end{align}
which results directly from Eqs. (\ref{zetarec1})-(\ref{zeta0k}).
Let us note that some of the indices of $U_{l\,n_1n_2}^{m_1m_2}$ must be increased only few times at most. In fact, the
maximal value of $6$ for the indices $n_1$, $n_2$ is sufficient to cover the whole known periodic table. Moreover, in
accurate calculations with Slater-type orbitals for light systems \cite{lesiuk15} one typically uses even-tempered
sequences of functions with $n=l+1$. This reduces the necessary values of $n_1$, $n_2$ to $1$. A similar observation is
valid for the $W_{lmn}$ integrals defined in the previous section, Eq. (\ref{wlmn}).

Evaluation of the integrals (\ref{jn1n2}) and (\ref{kn1n2}) follows a very similar strategy as adopted previously. By
using the power series expansion of $i_n(z)$ one easily arrives at
\begin{align}
 \mathcal{J}_{n_1n_2}^l = \beta_1^{n_1} \sum_{k=0}^\infty \frac{\left(\beta_1^2/2\right)^k}{k!\,(2n_1+2k+1)!!}
 \,\mathcal{I}_{2k+l+n_1,n_2}\big(0,\beta_2,\gamma\big).
\end{align}
The corresponding expression involving the second pair of indices is obtained by using the symmetry relation
$\mathcal{J}_{n_1n_2}^l(\beta_1,\beta_2,\gamma)=\mathcal{J}_{n_2n_1}^l(\beta_2,\beta_1,\gamma)$.
Both these formulae are useful for small or moderate $\beta_1$ or $\beta_2$, but fail otherwise due to slow convergence
of the infinite series. In this case one needs the large $\beta_1$ or $\beta_2$ asymptotic expansion which can be
derived analogously as Eqs. (\ref{asymi1}) and (\ref{asymi2}).

Finally, evaluation of the second class of integrals $\mathcal{K}_{n_1n_2}^l$ relies on the explicit expression for the
modified Bessel functions, Eq. (S3). By inserting it twice into the definition (\ref{kn1n2}) and
rearranging one obtains
\begin{align}
\begin{split}
&\mathcal{K}_{n_1n_2}^l = \frac{\pi^2}{4\beta_1\beta_2} \sum_{k_1=0}^{n_1}\sum_{k_2=0}^{n_2}
\frac{(n_1+k_1)!}{(2\beta_1)^{k_1}\,k_1!\,(n_1-k_1)!} \\
&\times\frac{(n_2+k_2)!}{(2\beta_2)^{k_2}\,k_2!\,(n_2-k_2)!}
\,\mathcal{F}^{\,>}_{l-k_1-k_2-2}\big(\beta_1+\beta_2,\gamma\big).
\end{split}
\end{align}

\subsection{Spin-orbit potentials}
\label{subsec:ecpso}

The effective spin-orbit potentials are of very similar form as the scalar pseudopotentials. In fact, they differ only
due to presence of the angular momentum and spin operators, Eq. (\ref{ecpso1}). Additionally, there is no local part in
the spin-orbit pseudopotentials. After some manipulations one can show that the necessary matrix elements 
\begin{align}
 \label{ibcaspin1}
\begin{split}
 &I_{\mbox{\scriptsize bca}}^{\mbox{\scriptsize so}} = 
 \langle\chi_{n_bl_bm_b}(\textbf{r}_b;\beta_b)|\,\hat{U}_{\mbox{\scriptsize
 PP,so}}^c(r_{c})\,|\chi_{n_al_am_a}(\textbf{r} _a;\beta_a)\rangle,
\end{split}
\end{align}
can be rewritten without a loss of generality as
\begin{align}
\label{ibcaspin2}
\begin{split}
 I_{\mbox{\scriptsize bca}}^{\mbox{\scriptsize so}} &=  
 \sum_{l=0}^{L-1}\sum_{mm'=-l}^{+l} \frac{2\,\dot{\imath}^{-1}}{2l+1} \int_0^\infty dr_c\;r_c^2\;
 \Delta U_l^c(r_{c}) \\
 &\times \langle\chi_{n_bl_bm_b}(\textbf{r}_b;\beta_b)|lm\rangle_c \cdot
 \,_c\langle lm|\,\mathbf{l}_{c}\cdot\mathbf{s}\,|lm'\rangle_c \\
 &\times\,_c\langle lm'|\chi_{n_al_am_a}(\textbf{r}_a;\beta_a)\rangle.
\end{split}
\end{align}
To derive this expression one uses the fact that the projection operators $_c\langle lm|$ are idempotent and that they
commute with the spin-orbit operator. The imaginary unit has been added to make all matrix elements real as the orbital
angular momentum operator is, in general, complex valued. The only new objects present in
Eq. (\ref{ibcaspin2}) are matrix element of the angular momentum operator, $\langle
lm|\,\mathbf{l}\cdot\mathbf{s}\,|lm'\rangle$. Explicit expressions for these integrals can be derived with standard
algebra of the angular momentum (see Ref. \cite{arfken}, pp. 793).

Standard quantum chemistry packages compute all basic matrix elements over spatial orbitals and the spin component is
added later by proper construction of an approximate wavefunction. This is the approach we adopt here. The integrals
(\ref{ibcaspin1}) and (\ref{ibcaspin2}) are evaluated for all Cartesian components separately and stored for further
manipulations.

\section{Core polarisation matrix elements}
\label{sec:cpp}

In order to evaluate the core polarisation correction to the Hamiltonian, Eqs. (\ref{cpp1})-(\ref{cpp4}), only two
distinct matrix elements are necessary. They read
\begin{align}
 \label{ibcacpp1}
\begin{split}
 I_{\mbox{\scriptsize bca}}^{\mbox{\scriptsize CPP}(i)} = 
 \langle\chi_{n_bl_bm_b}(\textbf{r}_b;\beta_b)|\,\hat{V}_{\mbox{\scriptsize
 CPP}}^{(i)}(r_{c})\,|\chi_{n_al_am_a}(\textbf{r} _a;\beta_a)\rangle,
\end{split}
\end{align}
with $i=1,2$, and
\begin{align}
\label{v1cpp}
 &\hat{V}_{\mbox{\scriptsize CPP}}^{(1)}(r) = \sqrt{\frac{4\pi}{3}}\,\frac{Y_{1M}(\hat{r})}{r^2}\,
\big(1 - e^{-\delta r^2}\big)^n,\\
\label{v2cpp}
&\hat{V}_{\mbox{\scriptsize CPP}}^{(2)}(r) = \frac{1}{r^4}\,\big(1 - e^{-\delta r^2}\big)^{2n}.
\end{align}
The Gaussian factors in these definitions come from the adopted cutoff function, Eq. (\ref{cut}). Note that instead of
Cartesian coordinates in Eq. (\ref{v1cpp}) we use pure spherical components corresponding to $M=-1,0,+1$. The total
contribution to the Hamiltonian can be assembled by combining these matrix elements with geometric and molecular
data according to Eqs. (\ref{cpp1})-(\ref{cpp4}).

Starting with the atomic-type integrals, one can straightforwardly integrate over the angles in the spherical coordinate
system, giving after some rearrangements
\begin{align}
\begin{split}
 &I_{\mbox{\scriptsize aaa}}^{\mbox{\scriptsize CPP}(1)} = 
 (-1)^{m_a'} \sqrt{(2l_a+1)(2l_a'+1)} \wigner{l_a'}{1}{l_a}{0}{0}{0} \\
 &\times \wigner{l_a'}{1}{l_a}{-m_a'}{M}{m_a} \sum_{k=0}^n \binom{n}{k} (-1)^k\,
\mathcal{F}^{\,0}_{n_a+n_a'-2}\big(\beta_a+\beta_a',\delta\big),
\end{split}
\end{align}
provided that $m_a'=m_a+M$ (otherwise the result vanishes). The form of the expression for the matrix element involving
$\hat{V}_{\mbox{\scriptsize CPP}}^{(2)}(r)$ depends on the value of $n_a+n_a'$. It reads
\begin{align}
 I_{\mbox{\scriptsize aaa}}^{\mbox{\scriptsize CPP}(2)} = \sum_{k=0}^{2n} \binom{2n}{k}
 (-1)^k\,\mathcal{F}^{\,0}_{n_a+n_a'-4}\big(\beta_a+\beta_a',k\delta\big),
\end{align}
for $n_a+n_a'\geq 4$,
\begin{align}
\begin{split}
 I_{\mbox{\scriptsize aaa}}^{\mbox{\scriptsize CPP}(2)} &= 
 \big( \beta_a+\beta_a' \big) \sum_{k=0}^{2n} \binom{2n}{k} (-1)^k
 \,\mathcal{G}^{\,0}_{0}(\beta_a+\beta_a',k\delta\big) \\
 &+ 4\delta n \sum_{k=1}^{2n-1} \binom{2n-1}{k-1}(-1)^k \,\mathcal{G}^{\,0}_{1}(\beta_a+\beta_a',k\delta\big),
\end{split}
\end{align}
\begin{align}
\begin{split}
 &I_{\mbox{\scriptsize aaa}}^{\mbox{\scriptsize CPP}(2)} = -\big( \beta_a+\beta_a' \big) \times \mbox{the above} \\
 &- 4\delta n \sum_{k=1}^{2n}
 \binom{2n-1}{k-1}(-1)^k \,\mathcal{F}^{\,0}_{0}\big(\beta_a+\beta_a',k\delta\big),
\end{split}
\end{align}
for $n_a+n_a'= 3$ and $n_a+n_a'= 2$, respectively. Let us remind that the above matrix elements are nonzero if and only
if $l_a=l_a'$ and $m_a=m_a'$.

Passing to the two-centre matrix elements, we first note that calculation of $I_{\mbox{\scriptsize
baa}}^{\mbox{\scriptsize CPP}(1)}$ and $I_{\mbox{\scriptsize bab}}^{\mbox{\scriptsize CPP}(1)}$ is almost exactly the
same as for the local components of the spin-averaged potentials described in Sections \ref{subsec:ecpbaa} and
\ref{subsec:ecpbab}. Thus, there is no need to repeat the details of the derivation and we provide only a
short sketch for convenience of the reader. Considering $I_{\mbox{\scriptsize baa}}^{\mbox{\scriptsize CPP}(1)}$, the
major difference as compared with the derivation given in \ref{subsec:ecpbaa} is that two spherical harmonics placed on
the centre $a$ need to be coupled first. Next, translation of the STO from the centre $b$ to the centre $a$ enables to
integrate
over the angles and the Jacobian cancels the apparent $1/r^2$ singularity introduced by the potential (\ref{v1cpp}).
This allows to expand the Gaussian damping function with help of the binomial theorem and the final result is written as
a linear combination of the $W_{lmn}$ integrals defined by Eq. (\ref{wlmn}). A similar conclusion is found for the
$I_{\mbox{\scriptsize bab}}^{\mbox{\scriptsize CPP}(1)}$ integrals class. Two STOs present on the centre $a$ need to be
expanded into a linear combination of STOs giving an analogue of Eq. (\ref{babloc}). Once we translate the distribution
from centre $b$ to centre $a$ and integrate over the angles, the singularity vanishes and rest of the
derivation is straightforward. The final result can also be written in terms of the integrals (\ref{wlmn}).

Unfortunately, calculation of the matrix elements involving the potential $\hat{V}_{\mbox{\scriptsize CPP}}^{(2)}(r)$
is more involved. This is due to the fact that the apparent singularity is not automatically cancelled by the Jacobian
and thus the damping factor in Eq. (\ref{v2cpp}) cannot be expanded that easily. As a result, in addition to the
ordinary integrals introduced in Sec. \ref{subsec:ecpbaa} and \ref{subsec:ecpbab} one requires
\begin{align}
\label{defbigibar}
&\widetilde{\mathcal{I}}_{ln}^p\big(\alpha,\beta,\delta\big) = \int_0^1 dx\; x^l\, i_n(\beta x)\,e^{-\alpha x}\, 
\Big( 1- e^{-\gamma x^2} \Big)^p, \\
\label{defbigkbar}
&\widetilde{\mathcal{K}}_{ln}^p\big(\alpha,\beta,\delta\big) = \int_1^\infty dx\; x^l\, k_n(\beta x)\,e^{-\alpha x}\, 
\Big( 1- e^{-\gamma x^2} \Big)^p,
\end{align}
where the analogy with Eqs. (\ref{defbigi}) and (\ref{defbigk}) is obvious. However, the values of $l$ are
not restricted to nonnegative integers here since $l=-1,-2$ are also necessary. For the integrals
$\widetilde{\mathcal{K}}_{ln}^p$ this is not problematic because of the integration range. Only the integrals
$\widetilde{\mathcal{I}}_{-1,n}^p$, and $\widetilde{\mathcal{I}}_{-2,n}^p$ are troublesome. To bring them into a closed
form we introduce the following quantities
\begin{align}
\label{defmln}
&\mathcal{M}_{ln}\big(\alpha,\beta,\gamma\big) = \int_0^1 dx\; x^l\,\ln x\, i_n(\beta x)\, e^{-\alpha
x-\gamma x^2},
\end{align}
so that $\widetilde{\mathcal{I}}_{-1,n}^p$ and $\widetilde{\mathcal{I}}_{-2,n}^p$ can now be simplified by
integration by parts. This gives
\begin{align}
\begin{split}
\widetilde{\mathcal{I}}_{-1,n}^p&\big(\alpha,\beta,\delta\big) = \\
&-\frac{n\beta}{2n+1}\sum_{k=0}^p \binom {p}{k} (-1)^k \mathcal{M}_{0,n-1}\big(\alpha,\beta,k\delta\big) \\ 
&-\frac{\beta(n+1)}{2n+1}\sum_{k=0}^p \binom {p}{k} (-1)^k \mathcal{M}_{0,n+1}\big(\alpha,\beta,k\delta\big) \\
&+\alpha \sum_{k=0}^p \binom {p}{k} (-1)^k \mathcal{M}_{0,n}\big(\alpha,\beta,k\delta\big) \\
&+4p\delta \sum_{k=1}^p \binom {p-1}{k-1} (-1)^k \mathcal{M}_{1,n}\big(\alpha,\beta,k\delta\big)
\end{split}
\end{align}
and
\begin{align}
\begin{split}
\widetilde{\mathcal{I}}_{-2,n}^p &= 
 \frac{\beta n}{2n+1}\,\widetilde{\mathcal{I}}_{-1,n-1}^p
+\frac{\beta(n+1)}{2n+1}\, \widetilde{\mathcal{I}}_{-1,n+1}^p \\
&-\alpha\, \widetilde{\mathcal{I}}_{-1,n}^p
-4p\delta\, \widetilde{\mathcal{I}}_{0,n}^p
+4p\delta\, \widetilde{\mathcal{I}}_{0,n}^{p-1}
\end{split}
\end{align}
where the notation for the nonlinear parameters $\big(\alpha,\beta,\gamma\big)$ has been suppressed when it is clear
from the context. Finally, calculation of the integrals $\mathcal{M}_{ln}$ is reminiscent of the methods introduced in
Sec. \ref{subsec:ecpbaa}. For example, for small and moderate $\beta$
\begin{align}
\label{mlnpow}
\mathcal{M}_{ln}\big(\alpha,\beta,\gamma\big) = \beta^n \sum_{k=0}^\infty \frac{\Big(\half
\beta^2\Big)^k}{k!\,(2n+2k+1)!!}\,\mathcal{G}^{\,<}_{n+l+2k}\big(\alpha,\gamma\big).
\end{align}
This finalises the present section of the paper.

\section{Numerical examples}

Throughout the paper we use atomic units for calculated quantities unless explicitly stated otherwise. The approximate
conversion factors are $1\,a_0 = 0.52\,917\,$\AA{} for lengths (Bohr radius), $1\,\mbox{a.u.}=219\,474.63$ cm$^{-1}$
for energies, and $1\,\mbox{a.u.}=2.54\,158$ Debye (D) for dipole moments.

\subsection{Basis set optimisation}

While there are many families of pseudopotentials available in the literature, the same cannot be said
about the relevant Slater-type basis sets. Therefore, we performed optimisation of the
valence STOs basis sets for three elements - calcium, strontium and barium (Ca, Sr, Ba). The last known element of the
rare-earth metals (radium, Ra) is not considered here because it is highly radioactive and thus not enough confirmed
experimental data is available to constitute a comprehensive test case. For all elements we adopted the
Stuttgart-Dresden family of energy consistent pseudopotentials. The so-called small-core pseudopotentials (10 valence
electrons) are described in Ref. \cite{lim06} whilst the large-core counterparts (2 valence electrons) are
given in Ref. \cite{fuentealba85}.

In general, construction of the STOs basis sets for pseudopotential calculations is similar as in the recent paper
concerning the beryllium dimer \cite{lesiuk15}. Therefore, we shall not repeat the minutiae of the procedure and
illuminate only the
most important differences. First, instead of the conventional even-tempered stencil for the nonlinear parameters
(exponents) of each angular momentum we use the following extended scheme (well tempering)
\begin{align}
 \zeta_{il} = \alpha_l\cdot\beta_l^{\,i\,\left(1+\gamma_l i + \delta_l i^2\right)},
\end{align}
where $i=0,\ldots,n_l$, $l$ is the angular momentum and $\alpha_l$, $\beta_l$, $\gamma_l$, $\delta_l$ are variational
parameters optimised for each $l$. For $l>2$ we set $\delta_l=0$ to reduce the number of parameters. The second
difference is the choice of the target function - total atomic valence correlation energy, \emph{i.e.}, we do not freeze
any additional orbitals in the valence space. Let us mention that there are many similarities between the basis set
optimisation procedures in the all-electron systems and for the valence-only pseudopotential. However, the latter case
is much more technically challenging. This is mainly due to occurrence of numerous local minima and problematic
behaviour of the pseudoorbitals near the nucleus causing linear dependencies problem.

The basis sets optimised in the course of the present work are constructed according to the correlation consistency
principle \cite{dunning89}. They are abbreviated wtcc-$l$ (well-tempered correlation-consistent), where $l$ is the
highest angular momentum present in the basis set. For example, for the valence-only ten electron systems (small-core
pseudpotentials) the smallest basis set (wtcc-2) has composition $10s8p3d$ and the largest (wtcc-5) - $13s11p7d5f4g2h$.
This includes two sets of additional diffuse functions which were trained to maximise the atomic polarisability
calculated at the closed-shell Hartree-Fock level. All basis sets used in this work can be obtained from the authors
upon request.

\begin{table}[t]
\caption{Results of the calculations for the calcium atom (see the main text for technical details). The abbreviation
IP stands for first ionisation potential of the system. Small-core PP subtracts 10 electrons from
the system (ECP10MDF) whilst large-core PP $-$ 18 electrons (ECP18SDF). All values are given in wavenumbers, cm$^{-1}$.}
\label{tabexca}
\begin{ruledtabular}
\begin{tabular}{ccccc}
 state & large-core PP & \multicolumn{2}{c}{small-core PP} & exp$^{\mbox{\scriptsize a}}$ \\[0.6ex]
       & CCSD          & CCSD & CC3 & \\
\hline \\[-2.2ex]
$^3$P & 15097.0 & 15173.2 & 15195.3 & 15263.1 \\
$^3$D & 20941.1 & 20856.1 & 21299.6 & 20356.6 \\
$^1$D & 22216.8 & 22878.6 & 22859.0 & 21849.6 \\
$^1$P & 23429.8 & 24845.8 & 23879.6 & 23652.3 \\
$^3$S & 31651.2 & 31828.7 & 31545.5 & 31539.5 \\ 
$^1$S & 33411.0 & 33890.9 & 33336.9 & 33317.3 \\
\hline \\[-2.2ex]
IP & 49405.2 & 49821.9 & --- & 49305.9 \\
\end{tabular}
\begin{flushleft}\vspace{-0.2cm}
$^{\mbox{\scriptsize a}}${\small experimental values taken from Refs. \cite{sugar85,miyabe06}; the experimental values
for the triplet states deduced from the Land\'{e} rule}
\end{flushleft}
\end{ruledtabular}
\end{table}

\begin{table}[t]
\caption{Results of the calculations for the strontium atom (see the main text for technical details). The abbreviation
IP stands for first ionisation potential of the system. Small-core PP subtracts 28 electrons from
the system (ECP28MDF) whilst large-core PP $-$ 36 electrons (ECP36SDF). All values are given in wavenumbers, cm$^{-1}$.}
\label{tabexsr}
\begin{ruledtabular}
\begin{tabular}{ccccc}
 state & large-core PP & \multicolumn{2}{c}{small-core PP} & exp$^{\mbox{\scriptsize a}}$ \\[0.6ex]
       & CCSD          & CCSD & CC3 & \\
\hline \\[-2.2ex]
$^3$P & 14579.9 & 14546.3 & 14597.2 & 14702.9 \\
$^3$D & 18442.2 & 18155.0 & 18393.7 & 18253.8 \\
$^1$D & 20380.4 & 20584.7 & 20411.1 & 20149.7 \\
$^1$P & 21451.1 & 22701.9 & 21797.5 & 21698.5 \\
$^3$S & 29201.4 & 29189.7 & 28939.3 & 29038.8 \\ 
$^1$S & 30634.4 & 31063.1 & 30508.6 & 30591.8 \\
\hline \\[-2.2ex]
IP & 46006.2 & 46284.4 & --- & 45932.2 \\
\end{tabular}
\begin{flushleft}\vspace{-0.2cm}
$^{\mbox{\scriptsize a}}${\small experimental values taken from Refs. \cite{sansonetti10,beigang82}; the experimental
values for the triplet states deduced from the Land\'{e} rule}
\end{flushleft}
\end{ruledtabular}
\end{table}

\subsection{Test results}

In order to check the accuracy of the new basis sets and correctness of the procedures given in this work we performed
extensive numerical tests. For each atom (Ca, Sr, Ba) we evaluated the first three excitation energies and the first
ionisation potential (IP). The results are given in Tables \ref{tabexca}-\ref{tabexba}. Additionally, in Tables
\ref{tabpolar} and \ref{taborb} we provide ground-state dipole polarisabilities (static) and outermost $ns$ valence
orbitals Hartree-Fock energies, respectively. All calculations were performed both with large- and small-core
pseudopotentials (2 and 10 valence electrons, respectively). In the case of the large-core pseudopotentials the
corresponding core polarisation potential was included by default. All valence two-electron calculations were performed
with the CCSD method \cite{purvis82} and its variants for the excited and ionised states (EOM, IP-EOM
\cite{geertsen89,stanton93,piecuch02}).

For the 10 electron systems (small-core pseudopotentials) the calculations are slightly more involved. For the excited
states we used the EOM-CC3 method \cite{koch97} as implemented in the code for excited state properties of Tucholska
\emph{et al} \cite{tucholska14,tucholska17} with all orbitals active. For the ionised states we used the IP-EOM2 method
\cite{gour05}, and the polarisabilities were evaluated at the CCSD and CCSD(T) \cite{raghavachari89} levels by using
two-point finite difference method with displacement of 10$^{-4}$ a.u. All calculations were performed with help of a
locally modified versions of the
\textsc{Gamess} \cite{gamess1,gamess2} and \textsc{AcesII} \cite{aces2} program packages, with an exception of the
computations at the CC3 level of theory where we used a program written by one of us (AMT). In all calculations
presented in this section the largest basis sets available in each case are used - wtcc-5 for the small-core
pseudopotentials and wtcc-3 for the large-core counterparts.

\begin{table}[t]
\caption{Results of the calculations for the barium atom (see the main text for technical details). The abbreviation
IP stands for first ionisation potential of the system. Small-core PP subtracts 46 electrons from
the system (ECP46MDF) whilst large-core PP $-$ 54 electrons (ECP54SDF). All values are given in wavenumbers, cm$^{-1}$.}
\label{tabexba}
\begin{ruledtabular}
\begin{tabular}{ccccc}
 state & large-core PP & \multicolumn{2}{c}{small-core PP} & exp$^{\mbox{\scriptsize a}}$ \\[0.6ex]
       & CCSD          & CCSD & CC3 &  \\
\hline \\[-2.2ex]
$^3$D & 9419.4  & 8923.7  & 9178.1  & 9357.8 \\
$^1$D & 11609.6 & 11653.5 & 11391.4 & 11395.5 \\
$^3$P & 12986.2 & 12823.6 & 12925.9 & 13085.5 \\
$^1$P & 17578.9 & 19527.3 & 18284.6 & 18060.3 \\
$^3$S & 26281.3 & 26269.3 & 26141.9 & 26160.3 \\ 
$^1$S & 27275.0 & ---$^{\mbox{\scriptsize b}}$ & ---$^{\mbox{\scriptsize b}}$ & 26757.3 \\
\hline \\[-2.2ex]
IP & 42156.4 & 42245.8 & --- & 42034.9 \\
\end{tabular}
\begin{flushleft}\vspace{-0.2cm}
$^{\mbox{\scriptsize a}}${\small experimental values taken from Refs. \cite{post85,karlsson99}; the experimental values
for the triplet states deduced from the Land\'{e} rule}\;\;
$^{\mbox{\scriptsize b}}${\small failed to converge}
\end{flushleft}
\end{ruledtabular}
\end{table}

\begin{table}[t]
\caption{Dipole polarisabilities of the ground state of the calcium, strontium and barium atoms. All values are given
in the atomic units.}
\label{tabpolar}
\begin{ruledtabular}
\begin{tabular}{cccc|c}
 atom & theory & large-core PP & small-core PP & exp \\[0.6ex]
\hline \\[-2.2ex]
\multirow{3}{*}{Ca} & HF      & 164.50 & 181.60 & \multirow{3}{*}{169$\pm$17$^{\mbox{\scriptsize a}}$} \\
                    & CCSD    & 170.38 & 159.14 &  \\
                    & CCSD(T) & ---    & 156.12 &  \\
\hline\\[-2.2ex]
\multirow{3}{*}{Sr} & HF      & 205.14 & 231.94 & \multirow{3}{*}{186$\pm$15$^{\mbox{\scriptsize b}}$} \\
                    & CCSD    & 221.48 & 203.16 &  \\
                    & CCSD(T) & ---    & 198.52 &  \\
\hline\\[-2.2ex]
\multirow{3}{*}{Ba} & HF      & 280.36 & 327.48 & \multirow{3}{*}{268$\pm$22$^{\mbox{\scriptsize c}}$} \\
                    & CCSD    & 323.48 & 284.70 &  \\
                    & CCSD(T) & ---    & 276.62 &  \\
\end{tabular}
\begin{flushleft}\vspace{-0.2cm}
$^{\mbox{\scriptsize a}}${Refs. \cite{schwartz74,miller76}}\;\;$^{\mbox{\scriptsize b}}${Ref.
\cite{miller02}}\;\;$^{\mbox{\scriptsize c}}${Ref. \cite{miller76}}\\
\end{flushleft}
\end{ruledtabular}
\end{table}

Let us begin the analysis with the atomic excitation spectra and consider the strontium atom as an example. The overall
picture is more-or-less the same for the remaining atoms and we shall comment on the differences further in the text.
One can see that both the small-core and large-core pseudopotentials give a very good agreement with the experimental
data.
However, the small-core pseudopotential combined with the CC3 method performs better, as could have been expected. The
average deviation from the experimental data is around 0.6\% for the small-core, and 0.9\% for the large-core
potentials. One can safely say that the ECP-MDF/CC3 level of theory is very reliable. On average, excitation energies
are expected to be less than 1\% away from the experimental data. Additionally, no significant increase of the error is
observed for any particular spatial symmetry or spin state. This suggests that the new basis sets have no inherent bias,
which is a desirable feature in a molecular work. 

Excitation energies for barium are in only slightly worse agreement with the experiment than in the case of strontium. 
The average error is around 0.9\% for the small-core and 1.7\% for the large-core pseudopotentials. Unfortunately, we
observe a significant error for the $^3$D and $^1$D states of calcium with both pseudopotentials. This behaviour is
surprising because the remaining excitation energies are in a good agreement with the experiment. Therefore, our first
suspicion was that $^3$D and $^1$D states are highly diffused and the basis set is not saturated well enough. However,
we found that further extension of the basis set changed the results by less that 100 cm$^{-1}$ which is not enough to
explain the discrepancy. As a result, we presume that this increase in the error is an inherent problem of the given
pseudopotentials. We note that in the original papers describing the pseudopotentials \cite{fuentealba85,lim06} errors
obtained for Ca were in fact significantly larger than for the other elements.

Let us also compare our results for the strontium atom with the values obtained by Skomorowski \emph{et al.}
\cite{skomorowski12b} In this work the same pseudopotential (ECP28MDF) was used in combination with a custom-made GTOs
basis set and the EOM-CC3 method. Both basis sets are roughly of the same size, so a fair comparison is possible.
Skomorowski \emph{et al.} \cite{skomorowski12b} give 14570.8 cm$^{-1}$ and 21764.3 cm$^{-1}$  for the nonrelativistic
$^3$P and $^1$P states, respectively. These results are very similar to the values given in Table \ref{tabexsr}; any
differences are probably accidental, suggesting that both basis sets are of a similar quality for the P states. However,
the situation is different for the D states. The authors of Ref. \cite{skomorowski12b} report 18668.8 cm$^{-1}$ for the
$^3$D state and 20650.3 cm$^{-1}$ for the $^1$D state. Clearly, errors with respect to the experimental values are much
larger than for the P states, and also by few hundreds cm$^{-1}$ larger than calculated with our basis sets (cf. Table
\ref{tabexsr}).

\begin{table}[t]
\caption{Outermost valence orbital energies calculated with the pseudopotentials compared with the reference
all-electron Dirac-Hartree-Fock (DHF). All values have their signs reversed and are given in the atomic units.}
\label{taborb}
\begin{ruledtabular}
\begin{tabular}{lcccc}
 atom & shell & large-core PP & small-core PP & all-electron DHF$^{\mbox{\scriptsize a}}$ \\[0.6ex]
\hline \\[-2.2ex]
Ca & 4$s$ & 0.2064 & 0.1967 & 0.1963 \\
Sr & 5$s$ & 0.1930 & 0.1813 & 0.1813 \\
Ba & 6$s$ & 0.1760 & 0.1630 & 0.1632 \\
\end{tabular}
\begin{flushleft}\vspace{-0.2cm}
$^{\mbox{\scriptsize a}}${\small taken from Ref. \cite{lim06}}
\end{flushleft}
\end{ruledtabular}
\end{table}

Next, we would like to check the quality of the basis sets for properties different than the atomic spectra. First, let
us consider the static dipole polarisabilities calculated with both families of pseudopotentials. The results are given
in
Table \ref{tabpolar}. The large-core pseudopotentials underperform considerably - the calculated values differ by more
than 10\% from the experimentally determined ones (and lie outside the corresponding error bars). The only exception is
the calcium atom, but this agreement is probably accidental. A completely different picture is found for the
small-core pseudopotentials. Here, calculated values are reasonably close to the experiment and lie within the given
error bars. We estimate the basis set error to be smaller than $1$ a.u. by observing the effect of additional diffuse
functions. Omission of the higher cluster operators brings an uncertainty of $1-2$ a.u. assuming that the results
converge geometrically with the excitation level. Therefore, one can expect that the theoretical limits are $2-3$ au.
below the values given in Table \ref{tabpolar}. This is still slightly above the experiment for Sr and Ba and somewhat
below for Ca. The remaining discrepancy might be a result of an inherent pseudopotential error or a systematic error in
the experimental data.

Lastly, we would like to consider the outermost valence $ns$ orbital energies calculated with the pseudopotentials and
compare them with all-electron Dirac-Hartree-Fock (DHF) values which we treat as a reference. Note that this quantity
is very important for chemical bonding phenomena and it is connected with some important descriptors such as the
electronegativity, \emph{etc.} The results are given in Table \ref{taborb}. Remarkably, the small core pseudopotentials
reproduce 3-4 significant digits for all atoms. The large-core counterparts are not that accurate and overestimate the
energy by 5\%-10\%. This alone allows that to predict that small-core pseudopotentials are expected to be much more
reliable in molecular studies.

\subsection{Results for diatomic systems}

\begin{table}[t]
\caption{Dissociation energy of the barium hydride (see the main text for technical details) calculated with
small-core pseudopotential (ECP46MDF). The abbreviations ``ae'' and ``fc'' stand for all-electron and frozen-core,
respectively. The quantity in the last column ($\Delta \mbox{fci}$) is the difference between the dissociation energies
calculated at the frozen-core FCI and CCSD(T) levels. The row denoted $\infty$ lists values extrapolated to the complete
basis set. All values are given in wavenumbers, cm$^{-1}$.}
\label{tabbah}
\begin{ruledtabular}
\begin{tabular}{ccccc}
 basis & ae-CCSD(T) & fc-CCSD(T) & $\Delta \mbox{fci}$ & total \\[0.6ex]
\hline \\[-2.2ex]
wtcc-2   & 13249.6 & 14239.9 & $+$4.6 & 13254.2 \\
wtcc-3   & 15701.2 & 15975.0 & $+$0.3 & 15701.5 \\
wtcc-4   & 16393.9 & 16355.8 & $-$1.2 & 16392.7 \\
wtcc-5   & 16645.7 & 16411.9 & $-$1.9 & 16643.8 \\
$\infty$ & 16903.9 & 16563.8 & $-$2.4 & {\bf 16901.5} \\
\hline \\[-2.2ex]
exp$^{\mbox{\scriptsize a}}$     & --- & --- & --- & $<$16350.0 \\
rev exp$^{\mbox{\scriptsize b}}$ & --- & --- & --- & $<$16910.6 \\
\end{tabular}
\begin{flushleft}\vspace{-0.2cm}
$^{\mbox{\scriptsize a}}${\small the original experimental value of Kopp \emph{et al.} \cite{kopp66}}\\
$^{\mbox{\scriptsize b}}${\small revision of the experimental value, Moore \emph{et al.} \cite{moore16}}
\end{flushleft}
\end{ruledtabular}
\end{table}

To keep length of the paper within reasonable limits we concentrate here on two molecules - strontium hydride and
barium hydride (SrH and BaH). Both of them have attracted a significant attention recently
\cite{lane15,abe10,aymar12,gao14,liu16}. We present results obtained with the more reliable small-core pseudopotentials
only. Analogous results for the large-core effective potentials can be obtained from the authors upon request.

For each of the molecules we evaluate the interaction energy ($D_e$) of the ground $X^2\Sigma^+$ state at the
experimentally determined geometry. We set the interatomic distance to $R=2.1461$ and $R=2.2319$ for SrH and
BaH, respectively, in accordance with the most recent experimental studies \cite{shayesteh04,ram13}. Additionally, we
evaluate the permanent dipole moment of both molecules and their vertical ionisation energy.

The procedure for evaluation of the aforementioned quantities is as follows. The interaction energy (\emph{i.e.}, the
well depth) is evaluated at the all-electron CCSD(T) level of theory by using the new basis sets, wtcc-$l$, with
$l=2,3,4,5$. Next, valence full triples correction is added, obtained at a difference between the frozen-core full CI
(FCI) and frozen-core CCSD(T) values. All results are extrapolated towards the complete basis set by using the ordinary
$L^{-3}$ formula. The ionisation energy is evaluated as a difference between the extrapolated CCSD(T) energies of the
molecule and the corresponding ion at a fixed geometry. Permanent dipole moments of the molecules are evaluated with
the finite field method by using displaced CCSD(T) energies. In contrast to the other quantities, basis set convergence
pattern for the dipole moment is not entirely regular. Thus, the extrapolation to the complete basis set has not been
attempted and we simply give values calculated with the largest available basis sets.

Let us begin the analysis with the interaction energy of the barium hydride. This quantity is interesting because of a
controversy connected with interpretation of the experimental data. The original experimental work of Kopp \emph{et
al.} \cite{kopp66} gives the value $D_e <16350.0$ cm$^{-1}$. However, in a recent paper of Moore \emph{et al.}
\cite{lane15} a significantly larger value has been obtained from \emph{ab initio} calculations, $D_e = 16895.12$
cm$^{-1}$. The discrepancy can be explained by assuming that the asymptote of one of the electronic states has been
incorrectly
identified. By selecting the correct Ba$(^3\mbox{D}_3)$ asymptote instead of Ba$(^3\mbox{D}_1)$, a revised experimental
value is obtained, $D_e <16910.6$ cm$^{-1}$. Our \emph{ab initio} results are given in Table \ref{tabbah} and the final
value,
$D_e = 16901.5$ cm$^{-1}$, supports the revision of the experimental data. The difference between the theoretical value
and the original experimental result ($\approx550$ cm$^{-1}$) is too large to be explained by the basis set error or the
pseudopotential error. Moreover, the agreement between our result and the value of Moore \emph{et al.} \cite{lane15} is
striking. Note that the $\Delta \mbox{fci}$ correction is very small for this molecule, of the order of few wavenumbers,
indicating that the CCSD(T) method works exceptionally well for this molecule.

Passing to the second molecule, strontium hydride, the corresponding results are given in Table \ref{tabsrh}.
Unfortunately, for this system we have no direct experimental results at our disposal. However, we can compare our
results with values reported in other theoretical papers. The most recent result of Liu \emph{et al.} \cite{liu16}
gives $D_e = 14114.6$ cm$^{-1}$, \emph{i.e.}, differing merely by 17 cm$^{-1}$ or about 0.1\%. Somewhat
older paper of Gao \emph{et al.} \cite{gao14} gives $D_e = 14259.8$ cm$^{-1}$ - a slightly larger deviation from our
value. However, let us note that a significantly smaller basis set was used in this work. Overall, it appears that the
newest theoretical values converge towards the most probable result around $D_e = 14100$ cm$^{-1}$. Parenthetically,
the values of the $\Delta \mbox{fci}$ correction are by an order of magnitude larger for SrH than for BaH, indicating
that the former possesses a much more pronounced multireference character. 

\begin{table}[t]
\caption{Dissociation energy of the strontium hydride (see the main text for technical details) calculated with
small-core pseudopotential (ECP28MDF). The abbreviations ``ae'' and ``fc'' stand for all-electron and frozen-core,
respectively. The quantity in the last column ($\Delta \mbox{fci}$) is the difference between the dissociation energies
calculated at the frozen-core FCI and CCSD(T) levels. The row denoted $\infty$ lists values extrapolated to the complete
basis set.
All values are given in wavenumbers, cm$^{-1}$.}
\label{tabsrh}
\begin{ruledtabular}
\begin{tabular}{ccccc}
 basis & ae-CCSD(T) & fc-CCSD(T) & $\Delta \mbox{fci}$ & total \\[0.6ex]
\hline \\[-2.2ex]
wtcc-2   & 12157.2 & 13221.7 & $+$28.1 & 12185.3 \\
wtcc-3   & 13561.4 & 14280.3 & $+$31.0 & 13592.4 \\
wtcc-4   & 13881.9 & 14428.6 & $+$29.2 & 13911.1 \\
wtcc-5   & 13982.5 & 14474.3 & $+$28.7 & 14011.2 \\
$\infty$ & 14103.9 & 14530.6 & $+$28.0 & {\bf 14131.9} \\
\hline \\[-2.2ex]
other theor. & --- & --- & --- & 14259.8$^{\mbox{\scriptsize a}}$ \\
             & --- & --- & --- & 14114.6$^{\mbox{\scriptsize b}}$ \\
\end{tabular}
\begin{flushleft}\vspace{-0.2cm}
$^{\mbox{\scriptsize a}}${\small Ref. \cite{gao14}}\;
$^{\mbox{\scriptsize b}}${\small Ref. \cite{liu16}}
\end{flushleft}
\end{ruledtabular}
\end{table}

\begin{table}[t]
\caption{Molecular properties of strontium and barium hydrides calculated with the small-core pseudopotentials. The
abbreviations $\mu$ and IP stand for the absolute values of the permanent electronic dipole moment and the (vertical)
ionisation potential. IPs and dipole moments are given in units of wavenumbers (cm$^{-1}$) and Debyes (D),
respectively.}
\label{tabdip}
\begin{ruledtabular}
\begin{tabular}{ccccc}
 & \multicolumn{2}{c}{SrH} & \multicolumn{2}{c}{BaH} \\
 & ae-CCSD & ae-CCSD(T) & ae-CCSD & ae-CCSD(T) \\[0.6ex]
\hline \\[-2.2ex]
 IP         & 42707.5 & 42917.6 & 38453.6 & 38791.6 \\
 $\mu$      & 13.49   & 13.53   & 14.30   & 14.38   \\[0.6ex]
\end{tabular}
\end{ruledtabular}
\end{table}

Finally, in Table \ref{tabdip} we present vertical ionisation potentials and permanent electronic dipole moments
calculated for both molecules. Unfortunately, these values are not directly comparable with any experimental data
available. Nonetheless, they can be used for comparison with other theoretical results, \emph{e.g.}, note
that the permanent dipole moments of SrH reported here are substantially larger than the values given by Gao \emph{et
al.} \cite{gao14}

To conclude this section we would like to comment on the computational efficiency of the procedures for calculation of
the pseudopotentials matrix elements. In all applications reported here we found these quantities to be much more
computationally expensive than the standard one-electron integrals, both in the atomic and diatomic systems. However,
this cost is still insignificant compared to the two-electron matrix elements of the electron-electron repulsion
operator. Therefore, calculations of the effective core potentials matrix elements do not constitute any significant
bottleneck within the present approach.

\section{Conclusions}

We have presented a general theory to evaluate matrix elements of effective core potentials in one-electron basis
set of Slater-type orbitals. As a rule, we have used the Barnett-Coulson translation method for STOs whenever possible.
It generates transparent formulae and all infinite summations truncate. As a result, the matrix
elements are reduced to relatively simple one-dimensional integrals. We have presented a scheme to evaluate them to a
very good precision.

Next, we have shown that the matrix elements of the spin-orbit pseudopotentials are reduced to the same basis quantities
as averaged effective potentials and only minor modifications are necessary to accomplish the calculations. Somewhat
larger changes are necessary to facilitate computations with the core polarisation potentials due to the apparent
singularities in the potential. Additional one-dimensional integrals with logarithmic singularities appear and we have
discussed their evaluation in details.

Finally, various numerical examples have been provided to verify the validity of the present approach. First, we have
shown a set of test results for the calcium, strontium and barium atoms, and compared the excitation energies, dipole
polarisabilities and valence orbital energies with reliable reference (exact or experimental) data. In all cases we have
found a very good agreement. Lastly, we have considered two molecular systems (strontium and barium hydrides) and
evaluated interaction energies, permanent dipole moments and ionisation energies; deviations from the available
experimental values have been found surprisingly small.

In this paper we have concentrated mainly on the diatomic molecules. However, the present approach can probably be
extended to an arbitrary polyatomic case with relative ease. This may be important for calculations in the spirit of
density functional theory \cite{adf01}, but also for general quantum chemical calculations for polyatomic systems in
the STOs basis in the face of recent improvements in many-centre STOs integrals technology \cite{jablczynska17}.

The code for evaluation of matrix elements of the effective core potentials in the STOs basis 
described in this paper has been incorporated in the \textsc{Ko\l os} program: \emph{general 
purpose ab initio program for electronic structure calculations with Slater-type orbitals, 
geminals, and Ko\l os-Wolniewicz functions}.

\begin{acknowledgments}
ML acknowledges the Polish Ministry of Science and Higher Education for the support through the project
\textit{``Diamentowy Grant''}, number DI2011 012041. AMT was supported by the Polish National Science Centre, project
number 2016/21/N/ST4/03734.
\end{acknowledgments}


\begin{thebibliography}{000}
\bibitem{foot1} Some authors prefer the term pseudopotential instead of the effective core potential; further in the
paper, those two terms shall be used interchangeably.
\end{thebibliography}
\end{document}

\title{{\Large Supplemental Material.} \\[0.5em] \noindent 
Combining Slater-type orbitals and effective core potentials}

\author{\sc Micha\l\ Lesiuk}
\email{e-mail: lesiuk@tiger.chem.uw.edu.pl}
\author{\sc Aleksandra M. Tucholska}
\author{\sc Robert Moszynski}
\affiliation{\sl Faculty of Chemistry, University of Warsaw\\
Pasteura 1, 02-093 Warsaw, Poland}
\date{\today}
\pacs{31.15.vn, 03.65.Ge, 02.30.Gp, 02.30.Hq}

\begin{abstract}
This document serves as a supplemental material for the publication \emph{``Combining Slater-type orbitals and
effective core potentials''}. It contains data or derivations which are of some importance but are not necessary for an
overall understanding of the manuscript. This includes detailed properties of the modified spherical Bessel functions
and numerical methods for calculation of various basic integrals defined in the main text.
\end{abstract}

\maketitle

\section{Selected properties of the modified spherical Bessel functions}

For convenience of the readers, we gather here all important properties of the modified spherical Bessel functions
which are used in the main text. Let us begin with the definitions
\begin{align}
 &i_n(z) = \sqrt{\frac{\pi}{2z}}\,I_{n+1/2}(z), \\
 &k_n(z) = \sqrt{\frac{\pi}{2z}}\,K_{n+1/2}(z),
\end{align}
where $I_n$ and $K_n$ are the usual modified Bessel functions. Both functions $i_n$, $k_n$ can be written in a
closed form involving only polynomials, inverse polynomials and exponential functions (\emph{i.e.}, they are
elementary). However, only the expression for the latter in practical
\begin{align}
 k_n(z) = \frac{\pi}{2}\,e^{-z} \sum_{k=0}^n \frac{(n+k)!}{2^k\,k!\,(n-k)!}\frac{1}{z^{k+1}}.
\end{align}
For $i_n$ series expansion around $z=0$ is typically used instead
\begin{align}
 i_n(z) = z^n \sum_{k=0}^\infty \frac{\left(\half z^2\right)^k}{k!\,(2n+2k+1)!!}.
\end{align}
Recursion relations which allow to increase the order of the modified spherical Bessel functions read
\begin{align}
i_{n-1}(z) - i_{n+1}(z) = \frac{2n+1}{z}\,i_n(z),\\
k_{n+1}(z) = \frac{2n+1}{z}\,k_n(z) + k_{n-1}(z).
\end{align}
Finally, both functions enjoy large-$z$ asymptotic expansions
\begin{align}
 &i_n(z) = \half\, e^z\, \sum_{k=0} (-1)^k \frac{a_{kn}}{z^{k+1}}, \\
 &k_n(z) = \frac{\pi}{2}\, e^{-z}\, \sum_{k=0}^\infty \frac{a_{kn}}{z^{k+1}}
\end{align}
where 
\begin{align}
 a_{kn} = \frac{1}{2^k k!} \prod_{i=1}^k (n+i)(n-i+1),
\end{align}
for $k>0$ and $a_{0n}=1$. Note that the above series are divergent and are valid only in the Poincare large $z$
asymptotic sense.

\section{Basic integrals with regular integrand - $\mathcal{F}^{\,0}_n\big(x,y\big)$ and related}

Let us start with the integrals $\mathcal{F}^{\,0}_n\big(x,y\big)$. By a simple exchange of variables they can be
rewritten as
\begin{align}
 \mathcal{F}^{\,0}_n\big(x,y\big) = \frac{x\,\Gamma(n+1)}{\left(2\sqrt{y}\right)^{n+2}}\,\,
\mathcal{U}\left(\half n+1,\frac{3}{2},\frac{x^2}{4y}\right),
\end{align}
where $\mathcal{U}(a,b,z)$ is the Tricomi function. Calculation of this quantity has already been solved even in more
general cases and we refer to specialised articles on the topic [for example, N. M. Temme, Numer. Math. \textbf{41}, 63
(1983)]. For the integrals $\mathcal{F}^{\,>}_n\big(x,y\big)$ and $\mathcal{F}^{\,<}_n\big(x,y\big)$ such simple
expressions cannot be found. However, the following recursive relations are helpful
\begin{align}
\label{recup}
 &\mathcal{F}^{\,>}_{n+2}\big(x,y\big) = \frac{1}{2y}\Big[-x\,\mathcal{F}^{\,>}_{n+1}\big(x,y\big) +
(n+1)\,\mathcal{F}^{\,>}_n\big(x,y\big) + e^{-x-y}\Big],\\
\label{recdw}
 &\mathcal{F}^{\,<}_n\big(x,y\big) = \frac{1}{n+1}\Big[e^{-x-y}+x\,\mathcal{F}^{\,<}_{n+1}\big(x,y\big)
 +2y\,\mathcal{F}^{\,<}_{n+2}\big(x,y\big)\Big].
\end{align}
It is crucial that these recursions are carried out as they stand, \emph{i.e.}, upward for
$\mathcal{F}^{\,>}_n\big(x,y\big)$ and downward for $\mathcal{F}^{\,<}_n\big(x,y\big)$. Otherwise, the numerical
stability of the algorithms is impacted negatively. To initiate the recursion (\ref{recup}) one starts with
\begin{align}
 \mathcal{F}^{\,>}_0\big(x,y\big)=\sqrt{\frac{\pi}{4y}}\,e^{\frac{x^2}{4y}}\,\,
\mbox{erf}\left(\frac{x+2y}{2\sqrt{y}}\right),
\end{align}
where $\mbox{erf}$ is the error function, and analogously for $\mathcal{F}^{\,>}_1\big(x,y\big)$. On the other hand, in
Eq. (\ref{recdw}) one initiates the recursion with $\mathcal{F}^{\,<}_N\big(x,y\big)=1$ and
$\mathcal{F}^{\,<}_{N+1}\big(x,y\big)=0$ at some large $N$, much larger than the actual maximal $n$ desired. The
results of the recursion must be subsequently scaled to match the exact value at $n=0$ calculated according to
\begin{align}
 \mathcal{F}^{\,<}_0\big(x,y\big)=\sqrt{\frac{\pi}{4y}}\,e^{\frac{x^2}{4y}}\bigg[
\mbox{erf}\left(\frac{x+2y}{2\sqrt{y}}\right) - \mbox{erf}\left(\frac{x}{2\sqrt{y}}\right)\bigg].
\end{align}
To avoid any further numerical instabilities we used the quadruple
arithmetic precision for both recursions. This does not introduce any significant computational overhead and is
sufficient for the range of parameters practically required. The above formulae contain no singularities when
$x=0$, and thus can be used with no changes in this special case. The singularities appearing as $y=0$ are immaterial
because they would require a nonphysical matrix element.

\section{Basic integrals with singular integrand - $\mathcal{G}^{\,0}_n\big(x,y\big)$ and related}

Calculation of the integrals with a singular integrand is very similar to the previous case. Recursion relations for
these integrals read
\begin{align}
 &\mathcal{G}^{\,0}_n\big(x,y\big) = \frac{1}{n+1}\Big[x\,\mathcal{G}^{\,0}_{n+1}\big(x,y\big)
 + 2y\,\mathcal{G}^{\,0}_{n+2}\big(x,y\big) - \mathcal{F}^{\,0}_n\big(x,y\big)\Big], \\
 &\mathcal{G}^{\,<}_n\big(x,y\big) = \frac{1}{n+1}\Big[x\,\mathcal{G}^{\,<}_{n+1}\big(x,y\big)
 + 2y\,\mathcal{G}^{\,<}_{n+2}\big(x,y\big) - \mathcal{F}^{\,<}_n\big(x,y\big)\Big], \\
 &\mathcal{G}^{\,>}_{n+2}\big(x,y\big) = \frac{1}{2y}\Big[(n+1)\,\mathcal{G}^{\,>}_{n}\big(x,y\big)
 - x\,\mathcal{G}^{\,>}_{n+1}\big(x,y\big) + e^{-x-y} + \mathcal{F}^{\,>}_{n}\big(x,y\big) \Big].
\end{align}
All above formulae where obtained by integration by parts and subsequent rearrangement. Again, these recursions need to
be carried out as written, \emph{i.e.}, downward in the case of the first two and upward in the third case. The
following integrals are necessary to complete the evaluation: $\mathcal{G}^{\,0}_0(x,y\big)$,
$\mathcal{G}^{\,<}_0(x,y\big)$, $\mathcal{G}^{\,>}_0(x,y\big)$. Calculation of an integral very closely related to them
has been discussed in detail recently [Silkowski \emph{et al.}, J. Chem. Phys. \textbf{142}, 124102 (2015)]. The
above three integrals can be solved by a rather straightforward generalisation of these techniques.